\documentclass[lettersize,journal]{IEEEtran}

\usepackage{amsmath,amsfonts}
\usepackage{algorithm}
\usepackage{algorithmic}
\usepackage{array}
\usepackage[caption=false,font=normalsize,labelfont=sf,textfont=sf]{subfig}
\usepackage{textcomp}
\usepackage{stfloats}
\usepackage{url}
\usepackage{verbatim}
\usepackage{graphicx}
\usepackage{newunicodechar}
\newunicodechar{−}{-}
\usepackage{orcidlink}
\usepackage{cite}
\usepackage[table,xcdraw]{xcolor}
\usepackage{multirow}

\begin{document}

\title{\textbf{CORVET}: A \textbf{CO}RDIC-Powered, \textbf{R}esource-Frugal Mixed-Precision \textbf{V}ector Processing \textbf{E}ngine for High-\textbf{T}hroughput AIoT applications}

\author{Sonu Kumar\orcidlink{0009-0000-4008-7153},
Mohd Faisal Khan\orcidlink{0009-0008-2235-8341},
Mukul Lokhande\orcidlink{0009-0001-8903-5159}, \IEEEmembership{Member, IEEE},\\ and
Santosh Kumar Vishvakarma\orcidlink{0000-0003-4223-0077}, \IEEEmembership{Senior Member, IEEE}
\thanks{Sonu Kumar, Mohd Faisal Khan and Santosh Kumar Vishvakarma acknowledge DST INSPIRE fellowship and MeitY/SMDP-C2S for ASIC tool support.
Sonu Kumar is with the Centre for Advanced Electronics, IIT Indore.
Mohd Faisal Khan, Mukul Lokhande, and Santosh Kumar Vishvakarma are with the NSDCS Research Group, Dept. of Electrical Engineering at IIT Indore.
\textbf{Corresponding author:} Santosh K. Vishvakarma (skvishvakarma@iiti.ac.in).}
}
\maketitle

\begin{abstract}
This brief presents a runtime-adaptive, performance-enhanced vector engine featuring a low-resource, iterative CORDIC-based MAC unit for edge AI acceleration. The proposed design enables dynamic reconfiguration between approximate and accurate modes, exploiting the latency-accuracy trade-off for a wide range of workloads. Its resource-efficient approach further enables up to 4× throughput improvement within the same hardware resources by leveraging vectorised, time-multiplexed execution and flexible precision scaling. With a time-multiplexed multi-AF block and a lightweight pooling and normalisation unit, the proposed vector engine supports flexible precision (4/8/16-bit) and high MAC density. The ASIC implementation results show that each MAC stage can save up to 33\% of time and 21\% of power, with a 256-PE configuration that achieves higher compute density (4.83 TOPS/mm\textsuperscript{2}) and energy efficiency (11.67 TOPS/W) than previous state-of-the-art work. A detailed hardware-software co-design methodology for object detection and classification tasks on Pynq-Z2 is discussed to assess the proposed architecture, demonstrating a scalable, energy-efficient solution for edge AI applications.

\end{abstract}

\begin{IEEEkeywords}
CORDIC, multiply-accumulate (MAC),  Non-linear Activation Functions, deep learning accelerators, Internet of things, Reconfigurable Computing.
\end{IEEEkeywords}

\section{Introduction}
\IEEEPARstart{D}{eep} learning has become a foundational component of modern artificial intelligence (AI) systems at the Internet of things (IoT), enabling breakthroughs across computer vision, speech recognition, natural language processing, and autonomous systems. Contemporary workloads are dominated by Deep Neural Networks (DNNs), Vision Transformers (ViTs), and emerging large-scale models, whose inference and training pipelines are primarily composed of convolutional layers, fully connected (FC) or multi-layer perceptron (MLP) blocks, and attention mechanisms~\cite{ILM, QuantMAC, Flex-PE, LPRE, NEURIC, access_ratko}. Despite architectural diversity, workload characterisation studies consistently show that multiply-accumulate (MAC) operations account for approximately 90\% of total computation, while non-linear activation functions (NAFs) contribute an additional 2-5\%~\cite{Flex-PE, NEURIC}. Efficient execution of these operations is therefore critical for deploying AI models on resource-constrained edge platforms\cite{TC6_edge,IoT7}.

To address the stringent energy, area, and latency constraints of edge AI, prior work has explored several hardware optimisation techniques, including fixed-point quantization~\cite{Flex-PE, POLARON}, CORDIC-based arithmetic~\cite{NEURIC}, logarithmic approximation~\cite{LPRE, QForce-RL}, and truncation-based MAC units~\cite{QuantMAC,HYDRA_ICIIS25}. While these approaches achieve notable reductions in computational complexity and power consumption, they typically operate at a fixed approximation point. As a result, either irreversible accuracy degradation is incurred, or additional error-compensation mechanisms are required, which partially negate the energy benefits~\cite{QuantMAC, Flex-PE}. Furthermore, these designs often lack the flexibility to dynamically adjust approximation depth based on layer sensitivity or application requirements\cite{Park-JSSC23, luca-JSSC25}.

Recent edge-oriented deep learning accelerators focus on energy efficiency, reduced memory traffic, and architectural flexibility by leveraging efficient dataflows, reconfigurable systolic arrays, and pipeline-aware designs. Complementary approaches exploit computing-in-memory, quantised and binary neural networks, and tightly integrated SoC platforms to alleviate bandwidth and storage bottlenecks requirements \cite{RAMAN, XRNPE_VLSID26}\cite{jaiswal2024quantized}. In parallel, recent studies have highlighted a structural inefficiency in many deep learning accelerators: the disproportionate allocation of hardware resources to activation-function units. Although AFs constitute a small fraction of total operations, they are frequently implemented using dedicated hardware blocks that remain idle for a significant portion of execution. Prior work reports up to 84\% idle cycles in NAF hardware for layer-reused architectures~\cite{HYDRA_ICIIS25}, while large-scale commercial accelerators such as Google TPUv4 allocate nearly 20–25\% of chip area to activation-related logic~\cite{Flex-PE}. This imbalance results in substantial dark silicon, limiting overall energy efficiency and scalability.

\begin{table*}[!t]
\caption{SoTA Design approaches and comparison of respective design features in AI workloads}
\label{tab:sota-comp}
\renewcommand{\arraystretch}{1.35}
\resizebox{\textwidth}{!}{%
\begin{tabular}{|c|c|c|c|c|c|c|c|c|}
\hline
\textbf{Design} & \textbf{Baseline} & \textbf{ICIIS'25}\cite{HYDRA_ICIIS25} & \textbf{ICIIS'25}\cite{HYDRA_ICIIS25} & \textbf{IEEE Access'24}\cite{QuantMAC} & \textbf{TVLSI'25}\cite{Flex-PE} & \textbf{ISCAS'25}\cite{LPRE} & \textbf{ISVLSI'25}\cite{NEURIC} & \textbf{Proposed} \\ \hline
\textbf{Compute} & Pipe-CORDIC & Pipe-CORDIC & \multicolumn{1}{c|}{PWL} & Pipe-CORDIC & Pipe-CORDIC & Logarithmic Approx. & Iterative CORDIC & Iterative CORDIC \\ \hline
\textbf{Arch. Type} & Fully Parallel & Layer-Reused & NAF-Reused & NAF-Reused & Systolic Array & \begin{tabular}[c]{@{}c@{}}Time-multiplexed \\ Reconfigurable Array\end{tabular} & Layer-Reused & Vector Engine \\ \hline
\textbf{Scalability} & No & Yes & No & No & - & Yes & No & Yes \\ \hline
\textbf{Precision} & FxP-8 & FxP-8 & FxP-8 & FxP-8 & FxP-4/8/16/32 & Posit-8/16/32 & FxP-8 & FxP-4/8/16 \\ \hline
\textbf{Accuracy loss} & High & High & High & High & Medium & Low & Medium & Variable (Low) \\ \hline
\textbf{Design Overhead} & Area, St. Power & Area & Area, St. Power & Area, Power & Energy & Area, Complexity & Latency & \begin{tabular}[c]{@{}c@{}}Application-optimized \\ High-Throughput\end{tabular} \\ \hline
\textbf{NAF-Supported} & ReLU & ReLU & Sigmoid/Tanh & NA & \begin{tabular}[c]{@{}c@{}}Sigmoid, Softmax, \\ Tanh, ReLU\end{tabular} & \begin{tabular}[c]{@{}c@{}}Sigmoid, Tanh, \\ Softmax\end{tabular} & Sigmoid/Tanh & \begin{tabular}[c]{@{}c@{}}SoftMax, GELU, Sigmoid\\ Tanh, Swish, ReLU, and SELU\end{tabular} \\ \hline
\textbf{Applications} & ANN & ANN & ANN & DNN & DNN, Transformers & DNN & DNN & DNN, Transformers (MLP) \\ \hline
\end{tabular}}
\end{table*}

Table~\ref{tab:sota-comp} provides a comparative overview of state-of-the-art (SoTA) AI accelerator designs, highlighting key architectural choices, supported precision, scalability, and associated trade-offs. Existing CORDIC- and approximation-based designs predominantly employ pipelined or fixed-stage implementations, which constrain runtime flexibility and enforce static accuracy-latency operating points\cite{IoT1, IoT2, IoT3, IoT4, TC3_iot}. In contrast, recent layer-reused or time-multiplexed architectures improve utilization but still lack fine-grained control over numerical accuracy at the MAC level. Consequently, prior designs are not very flexible with respect to different layer characteristics and operating conditions. They often require predefined datapaths and static precision settings \cite{Tp, TC1_mixprec}. This rigidity leads to suboptimal trade-offs among accuracy, latency, and resource utilisation when deployed across heterogeneous deep learning workloads \cite{Anurag}, particularly for models that demand mixed-precision computation and flexible activation support.

This paper addresses these limitations by proposing a runtime-adaptive, CORDIC-accelerated vector engine that explicitly exposes MAC precision, approximation depth, and execution latency as configurable architectural parameters. Unlike prior fixed-approximation designs, the proposed approach enables seamless switching between approximate and accurate execution modes without structural modification or auxiliary correction logic. In addition, a time-multiplexed multi-activation-function (multi-NAF) block is integrated to maximise hardware utilisation and significantly reduce dark silicon.

The key contributions of this work are summarised as follows:
\begin{enumerate}
    \item A low-resource, iterative CORDIC-based MAC unit with runtime-configurable accuracy-latency trade-offs, supporting both approximate and accurate execution modes.
    \item A scalable vector-engine architecture that amortises iterative MAC latency across parallel lanes, enabling higher throughput 4$\times$ without excessive area overhead.
    \item A high-utilisation, time-multiplexed multi-NAF block supporting a wide range of nonlinear functions with minimal additional hardware cost.
    \item A comprehensive evaluation spanning software emulation, FPGA prototyping, and 28\, nm ASIC synthesis, demonstrating system-level improvements on CNN and transformer-style workloads.
\end{enumerate}

The remainder of this paper is organised as follows. Section~II presents the proposed vector-engine architecture. Section~III details the circuit-level implementation of the iterative MAC and multi-NAF blocks. Section~IV describes the experimental methodology and evaluation framework. Section~V discusses FPGA, ASIC, and system-level results. Section~VI concludes the paper and outlines future research directions.

\section{Architecture Overview}

\begin{figure}[t]
    \centering
    \includegraphics[width=\columnwidth]{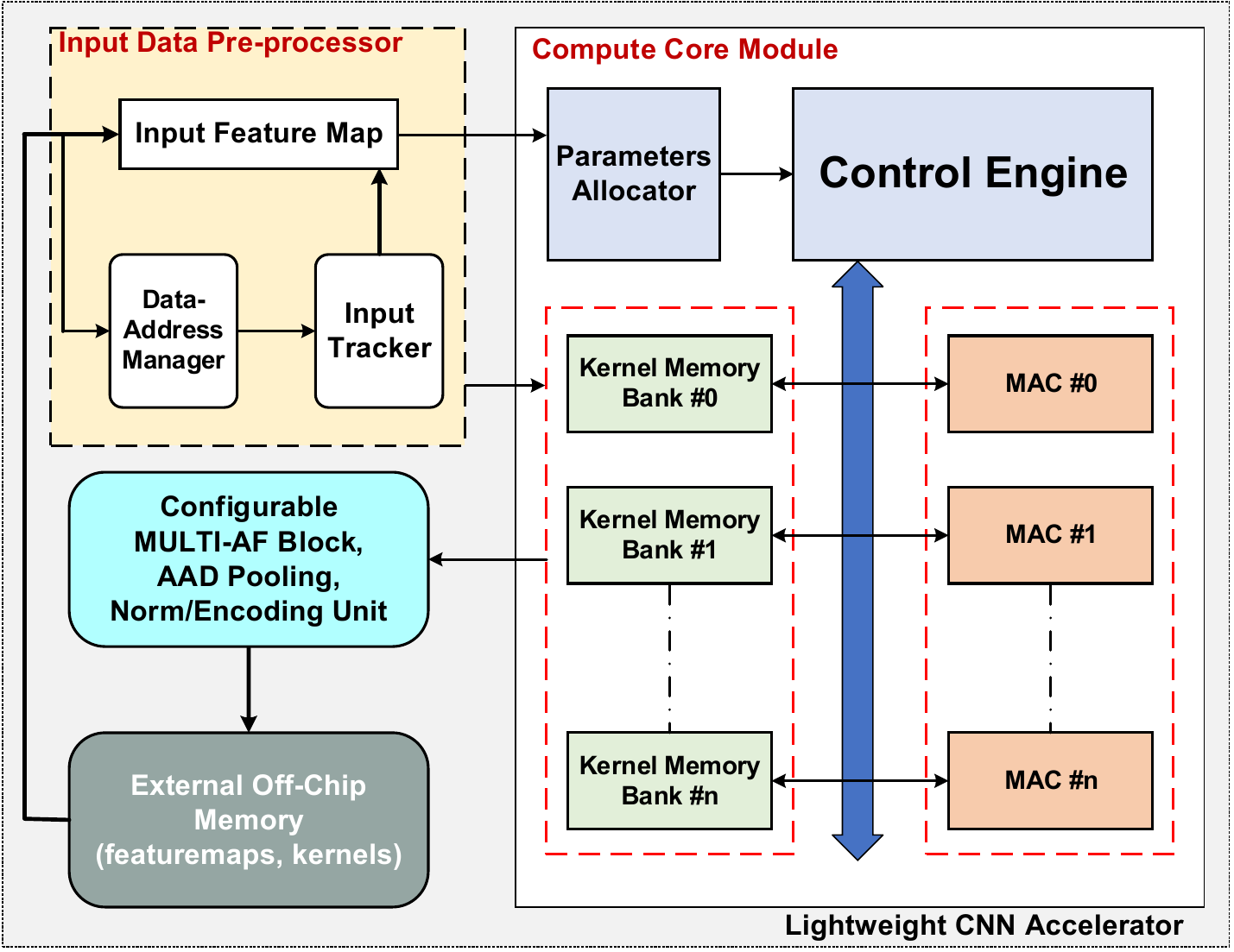}
    \caption{Block-level architecture of the proposed CORDIC-based vector engine integrated within a resource-efficient deep learning accelerator.}
    \label{fig:arch}
\end{figure}

Fig.~\ref{fig:arch} illustrates the top-level architecture of the proposed resource-efficient deep learning accelerator. The system is organised around a runtime-adaptive vector engine that serves as the primary compute core, supported by a lightweight control engine, data prefetcher, time-multiplexed multi-activation-function (multi-AF) block, pooling and normalisation units, and off-chip memory interfaces. The architecture is designed to maximise compute utilisation while enabling flexible trade-offs between accuracy and latency across diverse deep learning workloads.

\subsection{Vector Engine Organization}

The vector engine is composed of $N$ homogeneous processing elements (PEs), where $N$ is scalable from 64 to 256 depending on performance and area constraints. Each PE integrates a precision-adjustable, accuracy-configurable iterative CORDIC-based MAC unit, local register storage, and interface logic for data and control synchronisation. Unlike fully parallel systolic arrays, the proposed vector engine adopts a lane-based execution model, amortising the latency of iterative computations across multiple PEs to enable high throughput without requiring deeply pipelined or resource-intensive datapaths.

Two dedicated kernel memory banks, each organised as $(n$-bit $\times 32)$ entries, are employed to store input activations and weights, respectively. This dual-bank organisation enables continuous data feeding to the PEs while overlapping memory access with computation. The memory interface is designed to support flexible precision modes (4/8/16-bit) and runtime reconfiguration without stalling the compute pipeline.

\subsection{Runtime Accuracy and Precision Adaptation}

A key architectural feature of the proposed vector engine is its ability to dynamically adapt computation accuracy and latency at runtime. This is achieved by controlling the number of active CORDIC iterations within each MAC unit on a per-layer basis. Layers that are less sensitive to numerical error can be executed in an approximate mode with fewer iterations, thereby reducing latency and energy consumption. Conversely, accuracy-critical layers operate in a more accurate mode with additional iterations, incurring modest latency overhead while preserving numerical fidelity.

This runtime configurability is managed by the control engine through a set of configuration registers that specify precision mode, iteration count, and execution sequencing for each layer. Unlike prior approximation-based accelerators that rely on fixed hardware stages or static design-time tuning~\cite{Flex-PE, LPRE}, the proposed architecture enables fine-grained, layer-wise adaptation without requiring structural modification or auxiliary correction logic.

\subsection{Control Engine and Data Flow}

The control engine orchestrates vector-engine execution by managing instruction sequencing, memory addressing, and synchronisation across PEs. It comprises configuration registers for precision and iteration control, status registers for pipeline coordination, and a finite-state machine with datapath (FSMD) that governs layer execution. This lightweight control logic enables efficient coordination between compute, memory access, and activation processing while minimising control overhead.

\begin{figure}[t]
    \centering
    \includegraphics[width=\columnwidth]{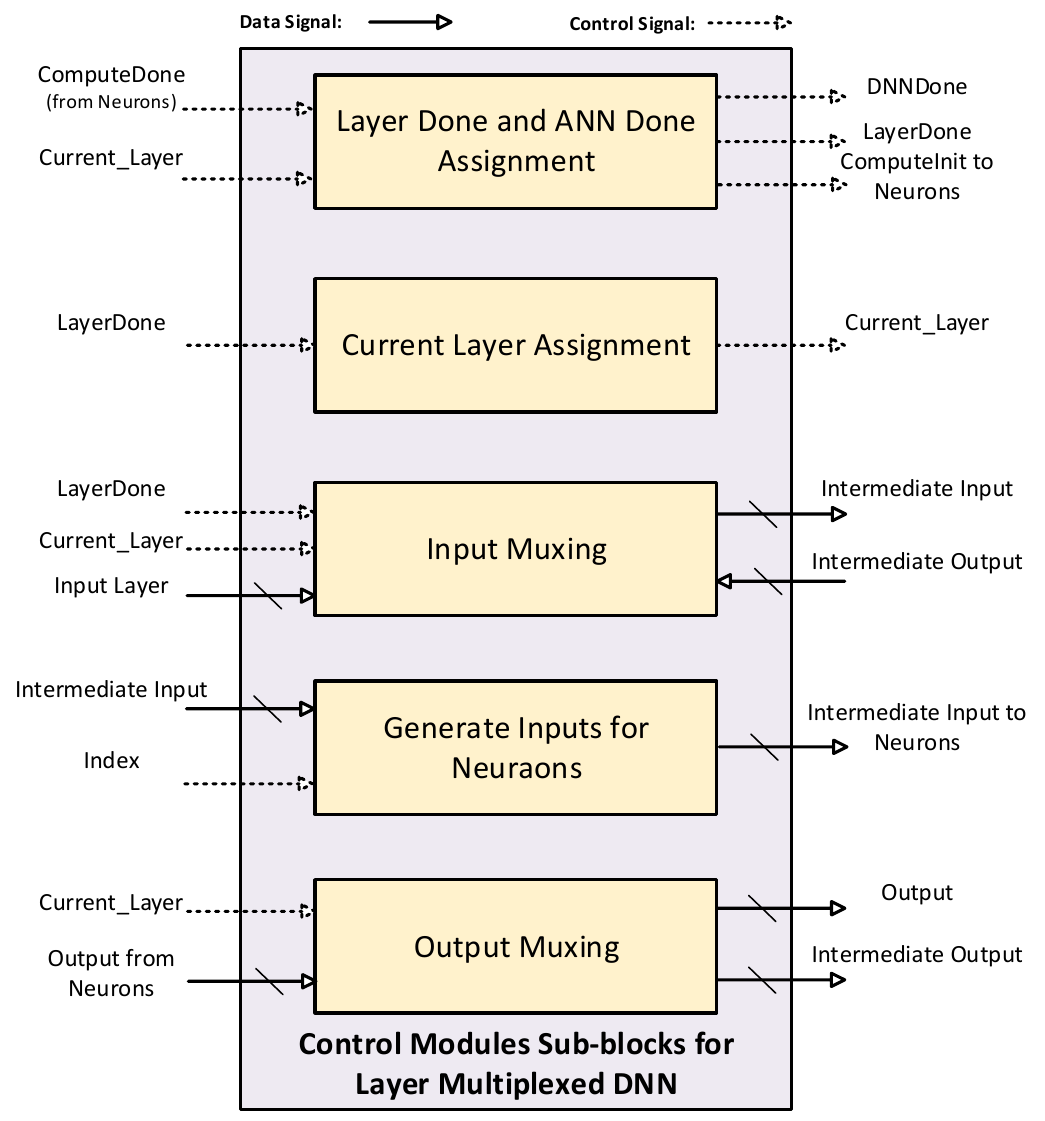}
    \caption{Control Engine for efficient reuse of data and control signals in a layer-multiplexed for reusing the same DNN architecture.}
    \label{fig:CE}
\end{figure}

A layer-multiplexed DNN requires a dedicated control unit composed of five functional sub-blocks, as illustrated in Fig.~\ref{fig:CE}. Each sub-block either monitors system status or produces control signals derived from that status. The control unit processes all status signals: LayerDone, DNNDone, CurrentLayer, ComputeInit, Index, and ComputeDone. The neuron processing elements produce Index and ComputeDone, two of the status signals, while the control module internally generates the remaining signals. The Index signal indicates which input to send next to the MAC unit by counting the number of MAC operations completed in the active layer. The ComputeDone signal indicates that neuron computation for the current layer has completed and that valid output data is available. When aggregated across all neuron units, this signal is referred to as ComputeDoneArray. Together, these status signals manage the data-path control necessary for the layer-multiplexed architecture to function properly.

The control module dynamically configures neuron activation and signal routing for layer-reused DNN computation. LayerDone and CurrentLayer are used to track progress, ComputeInit is used to selectively activate neurons per layer, and index-controlled input and output routes are used to multiplex intermediate data. This reduces dynamic power by enabling idle-unit deactivation and ensuring correct sequencing.

\begin{figure}[t]
    \centering
    \includegraphics[width=\columnwidth]{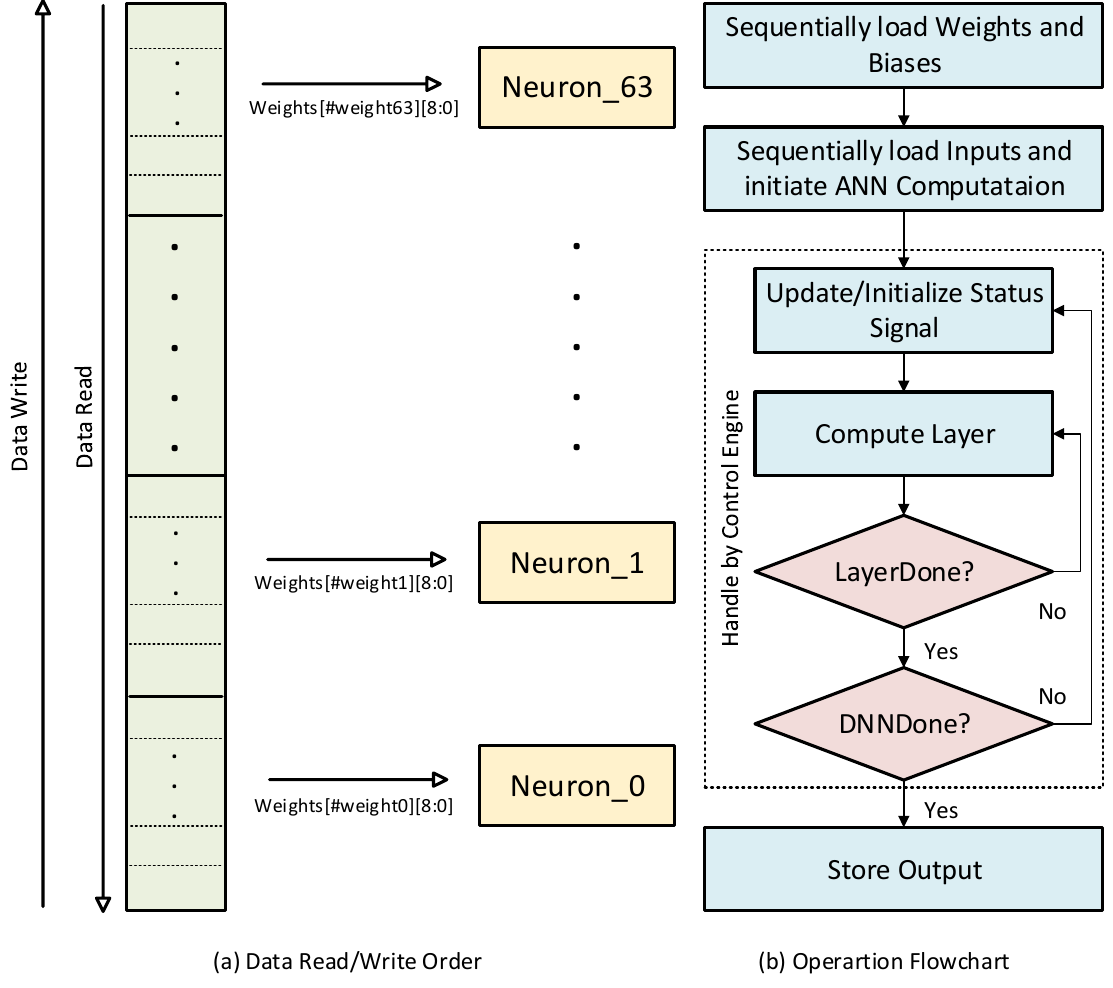}
    \caption{DNN accelerator data flow and order to initialise the loading data.}
    \label{fig:dataflow}
\end{figure}

As the weight memory is partitioned into 64 segments, each associated with a specific neuron processing unit, as depicted in Fig.~\ref{fig:dataflow}{(a)}. A key aspect of the parameter-loading mechanism is that the memory write sequence is the inverse of the read sequence. This organisation enables efficient access to weights and bias values with reduced interconnect delay. Consequently, parameters must be loaded using a Last-In–First-Out (LIFO) ordering for both weights and biases, as well as for input data. Data transfer follows a synchronous interface using a valid signal, denoted as \texttt{load\_param\_weight}, to indicate when weight values are written. The accelerator asserts a data-ready signal, DNNDone, once valid outputs are produced. Input values are accepted on each clock cycle when the valid signal is active. Upon completion, outputs from all ten neurons are generated simultaneously when DNNDone is asserted and are subsequently captured by the host software. The overall software-controlled execution sequence of the accelerator is illustrated in Fig.~\ref{fig:dataflow}{(b)}.

Data flow within the accelerator follows a streaming execution model. Input feature maps are fetched from off-chip memory through the data pre-fetcher and buffered locally before being broadcast to the vector engine. Partial sums generated by the PEs are forwarded either to subsequent MAC stages or to the activation function pipeline, depending on the layer configuration. This streaming approach minimises intermediate storage requirements and reduces memory bandwidth pressure, both of which are critical for energy-efficient edge deployment.

\subsection{Memory mapping}
\begin{figure}[t]
    \centering
    \includegraphics[width=\columnwidth]{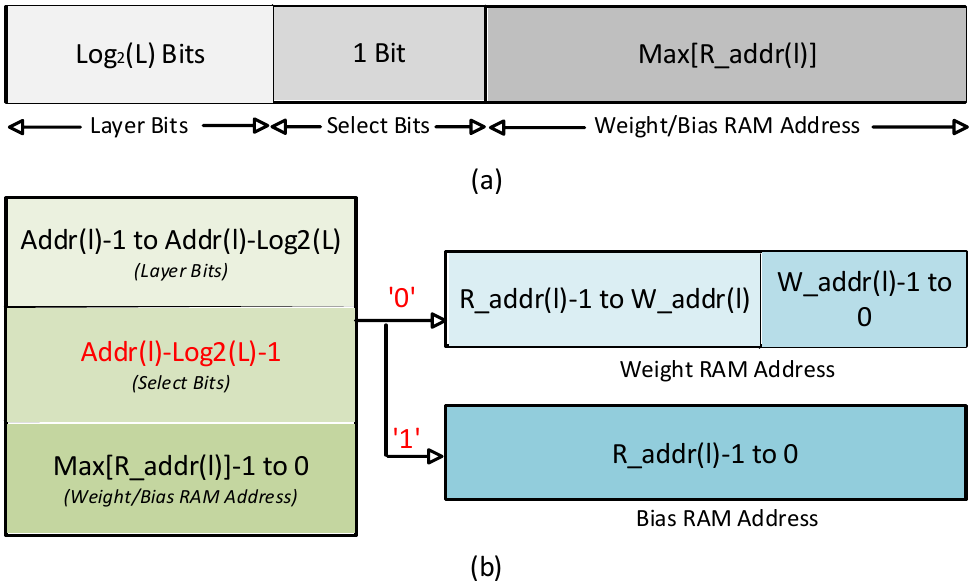}
    \caption{Memory mapping scheme for address bits that requires addressing weights and bias for the individual neurons.}
    \label{fig: memory map}
\end{figure}

The hardware architecture of fully connected neural networks must support scalable, adaptable setups, as the number of layers and neurons per layer varies across applications. As a result, the memory organisation for weights and biases must be adaptable and avoid allocating unused address locations, a limitation commonly encountered in fixed-addressing schemes. This is accomplished by using BRAM for on-chip parameter memory and FIFO buffers for temporary storage. An efficient addressing strategy for neuron-wise weight and bias access is illustrated in Fig.~\ref{fig: memory map} and is defined using the total number of layers 
L, the number of neurons in the lth layer N(l), and the number of inputs to that layer J(l). Since the number of neurons in one layer determines the number of inputs to the subsequent layer, these parameters satisfy
\begin{equation}
J(l+1) = N(l)
\end{equation}

Each parameter address consists of a layer identifier, a select bit indicating whether the accessed parameter is a weight or a bias, and a memory address field, as shown in Fig.~\ref{fig: memory map}{(a)}. The select bit distinguishes between weight and bias access,  while the most significant bits encode the layer index. The remaining bits represent either the neuron index (for bias) or the combined neuron and input index (for weight), as depicted in Fig.~\ref{fig: memory map}{(b)}. The required address length for weights and biases in a layer is therefore given by
\begin{equation}
R_{\text{addr}}(l) = \lceil \log_2 N(l) \rceil + \lceil \log_2 J(l) \rceil
\end{equation}
and that layer's overall address width turns into
\begin{equation}
\text{Addr}(l) = \lceil \log_2 L \rceil + 1 + R_{\text{addr}}(l)
\end{equation}
a fixed address width, which is chosen based on the maximum required across all levels, which is specified as
\begin{equation}
R_{\text{addr}} = \max_{l=1,2,\dots,L} 
\left\{ \lceil \log_2 N(l) \rceil + \lceil \log_2 J(l) \rceil \right\}
\end{equation}
and the length of the final uniform address is
\begin{equation}
\text{Addr} = \lceil \log_2 L \rceil + 1 + R_{\text{addr}}
\end{equation}
In addition to providing effective, conflict-free access to weight and bias memories for scaled DNN implementations, this technique enables consistent addressing.

\subsection{Time-Multiplexed Multi-Activation-Function Integration}

To address the underutilization of activation-function hardware observed in prior accelerators~\cite{HYDRA_ICIIS25, Flex-PE}, the proposed architecture integrates a time-multiplexed multi-AF block shared across all PEs. The multi-AF block supports a broad set of nonlinear functions, including Sigmoid, Tanh, SoftMax, GELU, Swish, ReLU, and SELU, using common CORDIC resources and mode-specific datapaths.

By multiplexing activation computation in time rather than dedicating separate hardware blocks, the architecture achieves high utilisation factors while incurring minimal area and power overhead. Activation execution is overlapped with vector-engine computation wherever possible, ensuring that the multi-AF block does not become a performance bottleneck despite being shared.

\subsection{Scalability and System Integration}

The proposed vector engine is designed for seamless scalability across edge and embedded platforms. By adjusting the number of PEs, memory bank sizes, and iteration depth, the architecture can be tailored to a wide range of performance and energy targets. Furthermore, the modular organisation of the vector engine, control logic, and peripheral units facilitates automated generation through a synthesizable hardware framework, enabling rapid design-space exploration and deployment.

Overall, the architecture combines runtime adaptability, high hardware utilisation, and scalable performance, forming a unified compute substrate that bridges the gap between fixed-approximation accelerators and fully accurate but resource-intensive designs.

\section{Circuit Implementation}

This section details the circuit-level design of the proposed iterative CORDIC-based MAC unit and the time-multiplexed multi-activation-function (multi-AF) block. The design objective is to achieve a balance between hardware efficiency, numerical accuracy, and runtime configurability while maintaining compatibility with standard deep learning workloads.

\subsection{Runtime-Adaptive Iterative CORDIC-Based MAC}

The proposed MAC unit is based on the unified CORDIC formulation originally introduced by Walther, which supports circular, linear, and hyperbolic computations using only shift, add/subtract, and multiplexing operations. Recent works such as ReCON~\cite{NEURIC} and Flex-PE~\cite{Flex-PE} have demonstrated the applicability of CORDIC arithmetic to deep learning operations, including MAC, Sigmoid, Tanh, and SoftMax~\cite{sumi_softmax}. However, these designs primarily employ pipelined or fixed-stage CORDIC architectures, which impose a static trade-off between accuracy and latency.

\begin{figure}[t]
    \centering
    \includegraphics[width=\columnwidth]{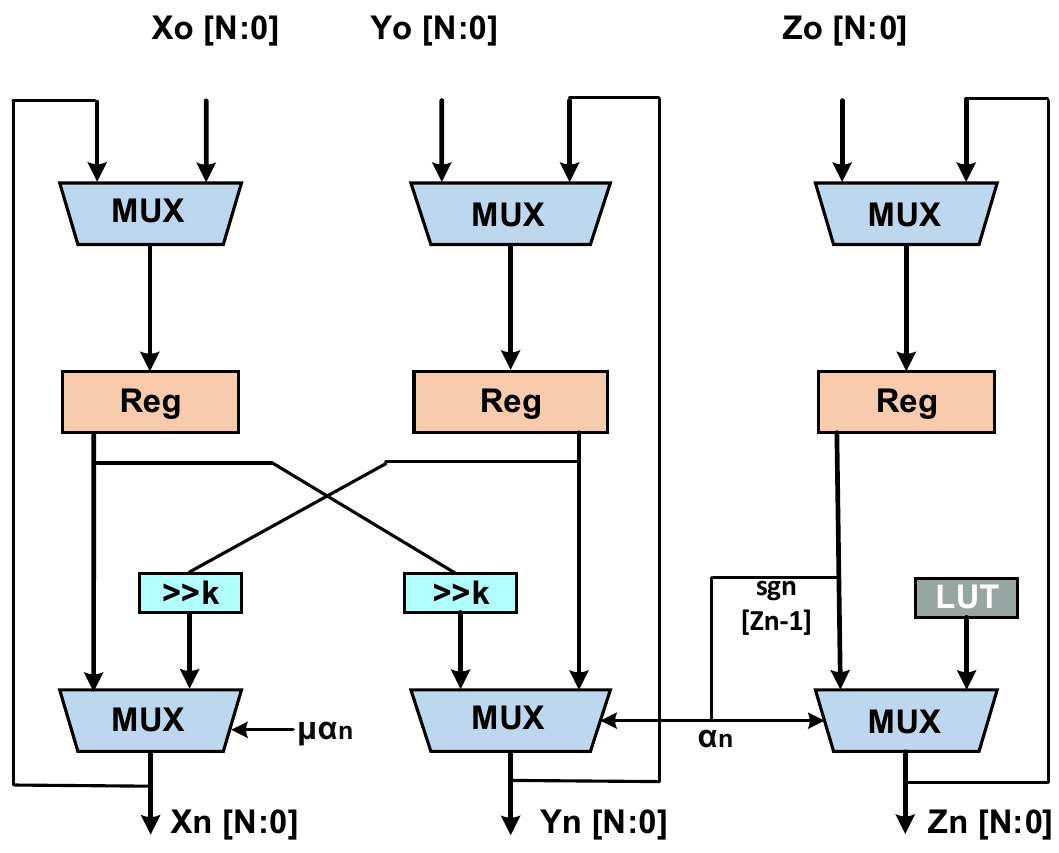}
    \caption{Iterative low-latency CORDIC-based MAC architecture with runtime-configurable iteration depth.}
    \label{fig:iter-cordic-mac}
\end{figure}

In contrast, the proposed MAC adopts an iterative CORDIC structure, as illustrated in Fig.~\ref{fig:iter-cordic-mac}, where the number of active iterations directly determines the approximation error and execution latency. This enables runtime switching between approximate and accurate execution modes without altering the hardware structure or introducing auxiliary correction logic.

The MAC unit supports both 8-bit and 16-bit fixed-point precision modes. In approximate mode, the MAC completes 8-bit and 16-bit operations in 4 and 7 clock cycles, respectively, incurring approximately 2\% accuracy degradation at the application level. In accurate mode, additional iterations are enabled, completing 8-bit and 16-bit operations in 5 and 9 cycles with less than 0.5\% accuracy loss. Additionally, it supports 4-bit modes with accurate 4-bit cycle operation.  These operating points are selected based on an accuracy-sensitivity heuristic~\cite{Flex-PE}, enabling layer-wise configuration based on numerical criticality.

From a circuit perspective, the iterative MAC minimises area and static power by reusing a single CORDIC datapath across iterations, rather than replicating pipeline stages. This design choice reduces the number of adders, shifters, and registers compared to pipelined alternatives, while still enabling high throughput at the vector-engine level through parallelism across multiple PEs.

\subsection{Latency Hiding Through Vector-Level Parallelism}

Although the iterative MAC incurs a multi-cycle latency per operation, this overhead is effectively hidden at the vector-engine level. Since multiple PEs operate concurrently on independent data elements, the increased per-MAC latency does not limit overall throughput for sufficiently large vector widths. This execution model distinguishes the proposed architecture from fully parallel or systolic-array designs, which require deeply pipelined datapaths to sustain throughput and therefore incur higher area and power overheads.

The ability to trade per-MAC latency for reduced hardware complexity is particularly advantageous for edge AI accelerators, where area and energy efficiency are often more critical than single-operation latency.

\begin{figure}[!t]
    \centering
    \includegraphics[width=1\linewidth]{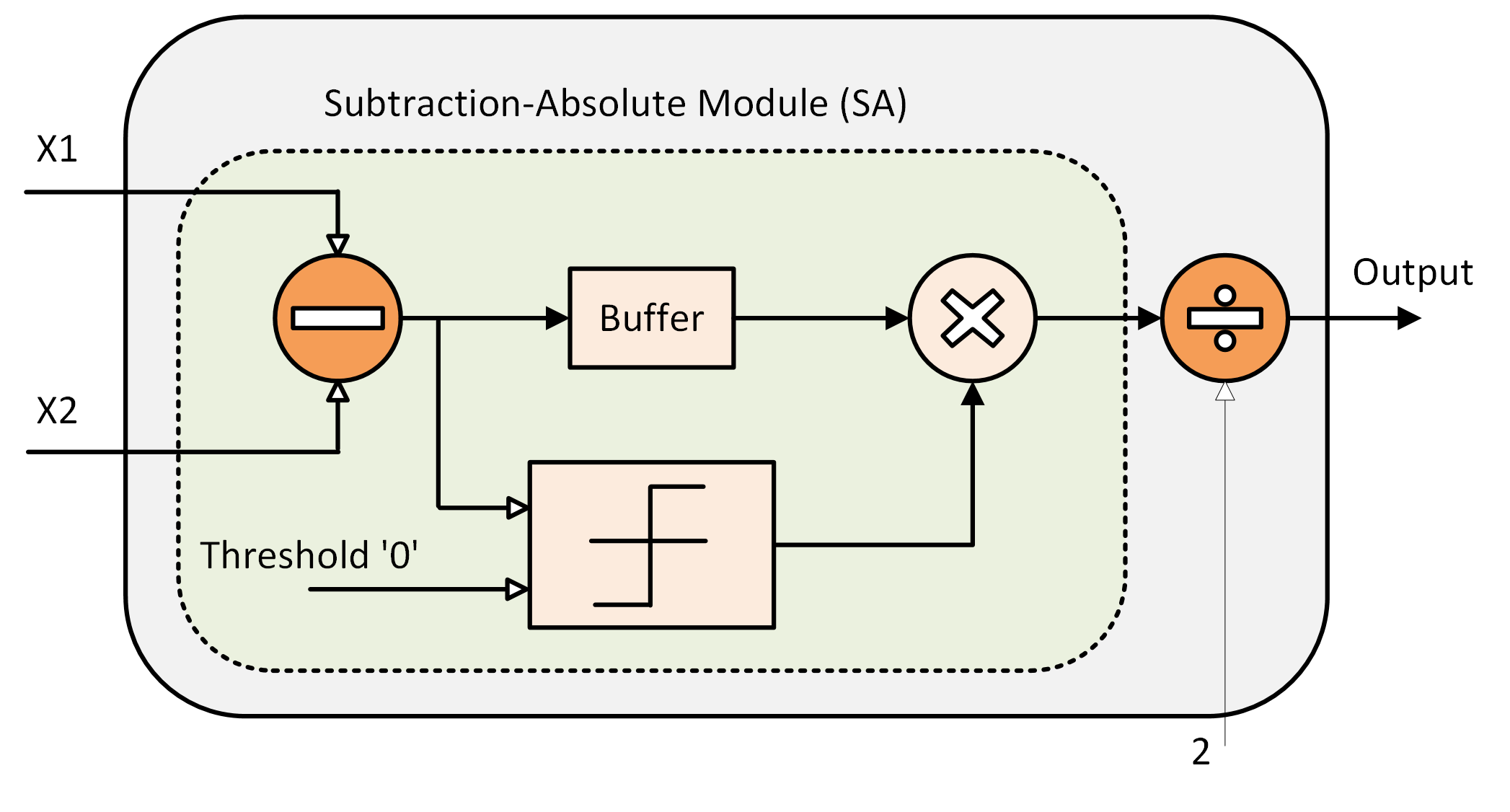}
    \caption{Hardware AAD module for two inputs.}
    \label{fig: AAD module}
\end{figure}

\begin{figure}[!t]
    \centering
    \includegraphics[width=1\linewidth]{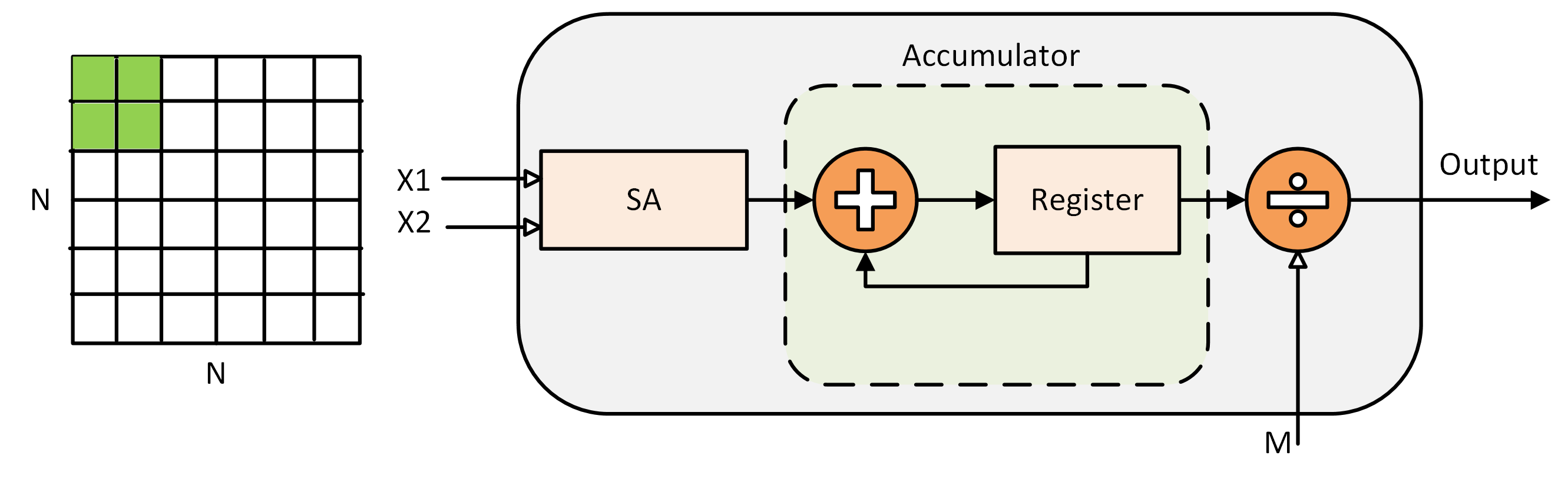}
    \caption{Hardware AAD module architecture based on sliding window}
    \label{fig:AAD module architecture}
\end{figure}

\subsection{Absolute Average Deviation (AAD) Pooling Block}

In addition to the MAC and activation units, the vector engine integrates peripheral components such as an Absolute Average Deviation (AAD) pooling unit~\cite{AAD-pool} and a normalisation block. The AAD pooling unit is selected due to its favourable accuracy characteristics for CORDIC-based computation, demonstrating a 0.5-1\% accuracy improvement over conventional pooling methods with lower computational complexity~\cite{Flex-PE, AAD-pool}.

Division, subtraction, and absolute value computation are the three primary steps of the hardware implementation of the two inputs in Average Absolute Deviation (AAD) unit as shown in Fig. \ref{fig: AAD module}. Initially, the two input values are fed into a subtractor to determine their difference. The subtraction result is then processed through two parallel paths. A comparator receives the result from one path and compares it to zero to identify the sign of the difference, returning either +1 or −1. To match its timing with the comparator output, the other channel passes the subtraction result through a buffer. These two outputs are multiplied, ensuring the final result is always non-negative regardless of the input order, effectively yielding the absolute deviation. This absolute deviation is then divided by two to obtain the final AAD output for the two-input case.

\begin{figure}[!t]
    \centering
    \includegraphics[width=1\linewidth]{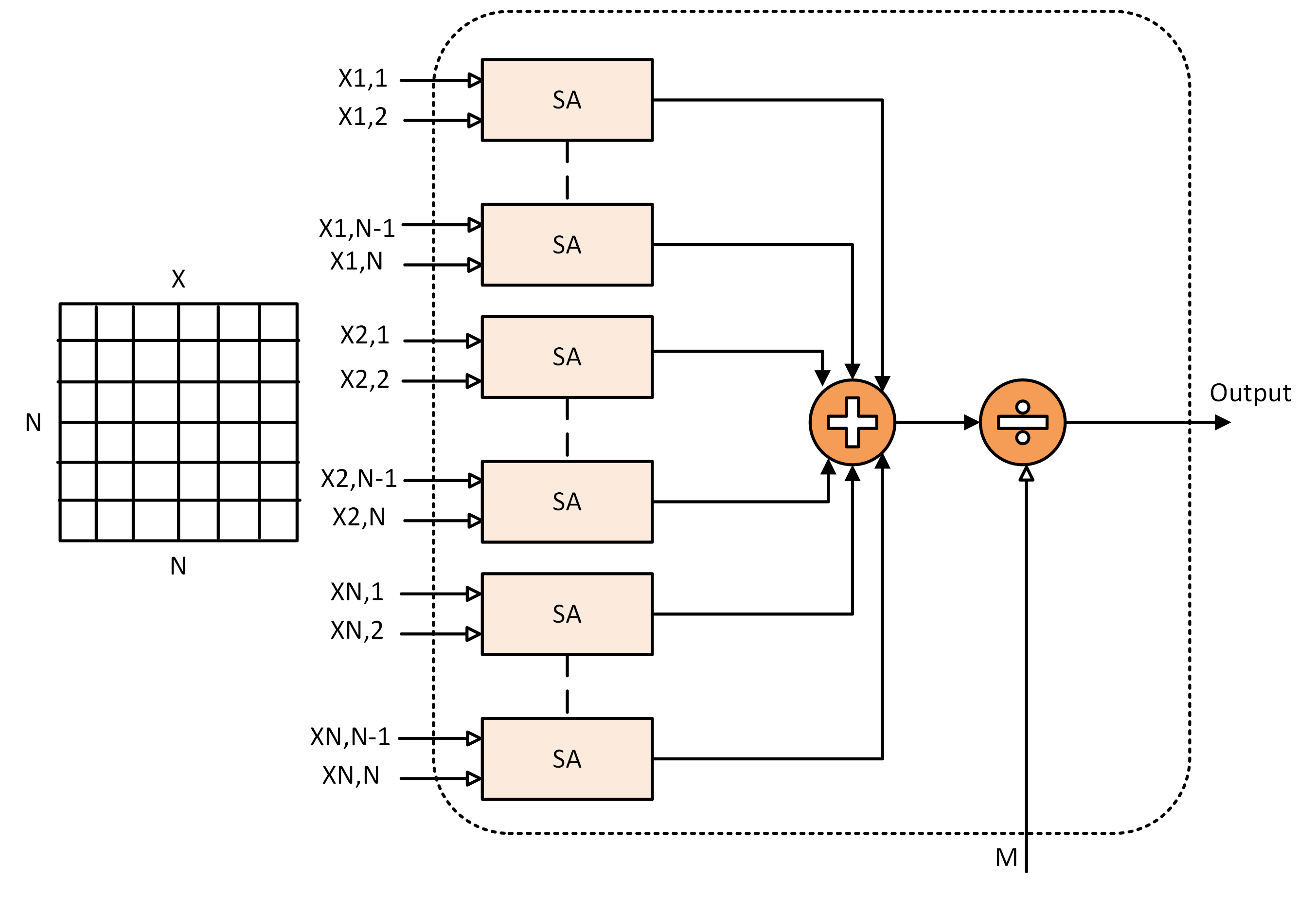}
    \caption{Hardware AAD module architecture based on parallel computation.}
    \label{fig:Hardware AAD}
\end{figure}

\begin{figure}[!t]
    \centering
    \includegraphics[width=1\linewidth]{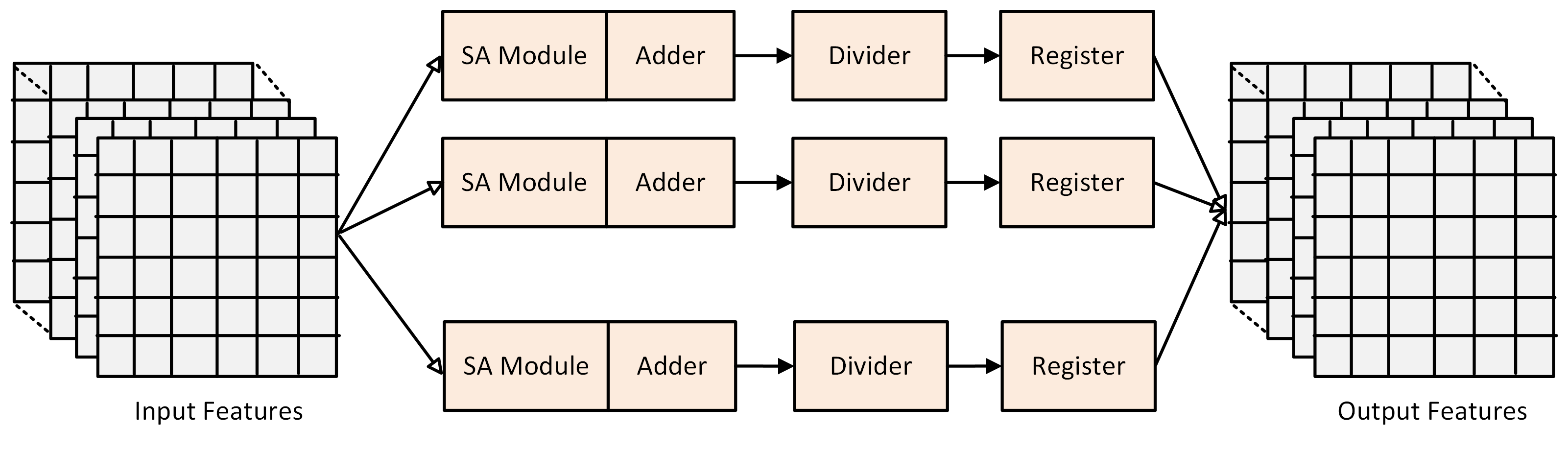}
    \caption{Multiple feature computations in parallel in hardware.}
    \label{fig:features_computation}
\end{figure}

For multi-input scenarios, multiple subtraction-absolute (SA) modules operate in parallel, each computing the absolute deviation between pairs of input values as shown in the Fig. \ref{fig:Hardware AAD}. The outputs of these SA modules are summed using an adder network, and the accumulated result is divided by the normalisation factor M = N (N−1) to produce the overall AAD value, which is carried out in parallel as shown in Fig. \ref{fig:features_computation}. A sliding window technique, in which a window moves over the input data with a specified stride and pooling size, is used to simplify the hardware. To reduce hardware complexity, a sliding window approach is adopted, with a window moving across the input data according to the defined stride and pooling size. Within each window, deviations between data points are computed, accumulated in registers, and normalised to produce the final AAD result efficiently as illustrated in Fig. \ref{fig:AAD module architecture}.

\subsection{Time-Multiplexed Multi-Activation-Function Block}

\begin{figure*}
    \centering
    \includegraphics[width=1\linewidth]{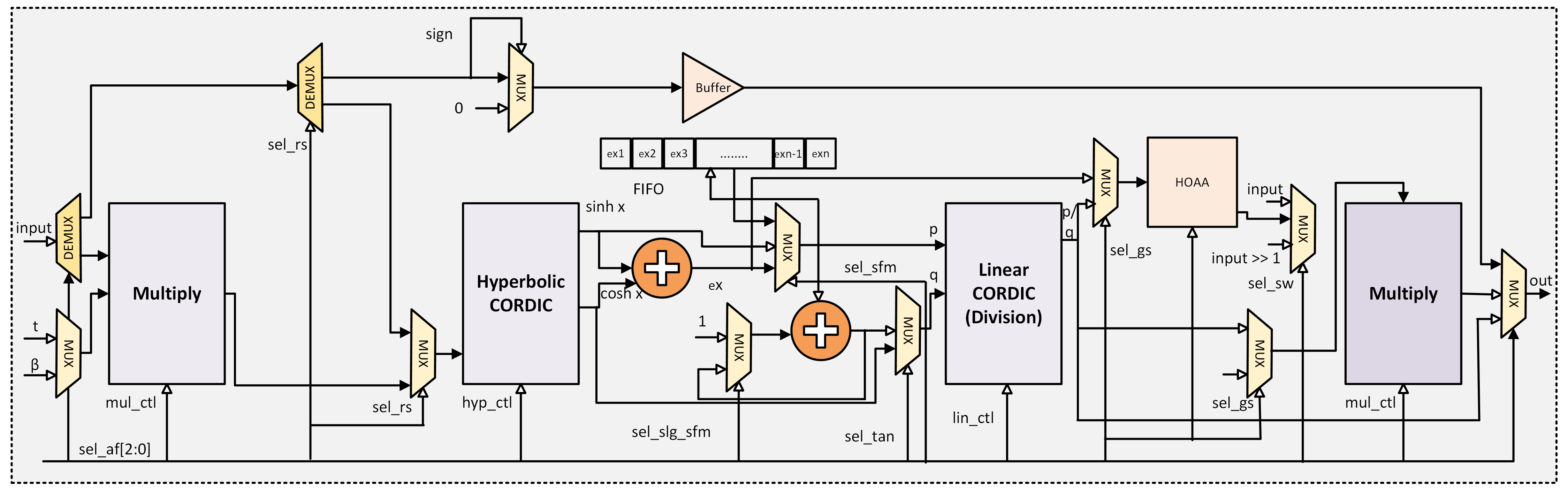}
    \caption{Time-multiplexed Activation function with integrated data flow and control signals.}
    \label{fig:af}
\end{figure*}

Activation functions represent a small fraction of total operations but often consume disproportionate hardware resources. To address this inefficiency, the proposed design integrates a time-multiplexed multi-AF block that reuses CORDIC hardware across multiple nonlinear functions as shown in Fig. \ref{fig:af}. The multi-AF block supports Sigmoid, Tanh, SoftMax, GELU, Swish, ReLU, and SELU, enabling compatibility with both CNN and transformer-style workloads.

The multi-AF block operates in two primary modes: a hyperbolic rotation (HR) mode for functions that require $\sinh$ and $\cosh$ computations, and a linear-division (LV) mode for functions that involve normalisation or exponential scaling. By selectively enabling only the required datapaths for a given function, the design achieves utilisation factors of up to 86\% in HR mode and approximately 72\% in LV mode.

Additional auxiliary logic includes a lightweight switching multiplexer for Sigmoid and Tanh selection, a ReLU bypass buffer, a FIFO for intermediate SoftMax storage, and two small multipliers\cite{TC4_mul} to support GELU computation. Collectively, these components incur less than 4\% additional area and power overhead while significantly improving overall hardware utilisation.

\subsection{Peripheral Support and Integration}

The proposed vector engine is integrated as a complete edge-AI processing subsystem comprising a lightweight control engine, on-chip memory banks, input pre-processing logic, and a host communication interface. The control engine uses configuration registers, status flags, and a finite-state machine to control memory addressing, instruction sequencing, and synchronisation, coordinating execution across the MAC array, activation block, and pooling units. Layer-adaptive execution is made easier by runtime control signals such as ComputeInit, LayerDone, and ComputeDone. This enables the reuse of hardware resources across multiple network levels, thereby guaranteeing proper data ordering. A data prefetcher retrieves input feature maps from external memory, buffers them locally, and then broadcasts them to processing components. Index-controlled multiplexing is used to transport intermediate outputs to later layers. To facilitate continuous data feeding and overlapping memory access with computation, parameter storage is managed via partitioned kernel memory banks that independently store activations and weights. The memory interface supports synchronous valid-data loading with a data-ready completion signal, allowing the host processor\cite{TC2_riscv} to capture final outputs without stalling the compute pipeline. All processing elements share a time-multiplexed multi-activation-function unit that uses common CORDIC resources to perform nonlinear operations. To prevent performance bottlenecks, its operations overlap with MAC computation. In addition, integrated pooling and normalisation blocks process partial sums before output generation, reducing intermediate storage and external memory traffic. The modular organisation of control, memory, and peripheral compute stages enables scalable deployment across FPGA and ASIC platforms. It supports efficient system-level integration with embedded processors through a lightweight interface, thereby transforming the vector engine from a standalone compute core into a deployable edge-AI accelerator.

\section{Experimental Methodology}

To ensure a rigorous, fair, and reproducible evaluation, the proposed vector engine is validated using a structured hardware-software co-design methodology spanning algorithmic emulation, RTL-level verification, FPGA prototyping, and ASIC synthesis. The evaluation framework is designed to isolate the impact of iterative CORDIC approximation while maintaining consistent experimental conditions across all comparisons.

\subsection{Software-Level Functional Emulation}

At the algorithmic level, an iso-functional software model of the proposed vector engine is developed in Python~3.0. The model emulates the vector engine's custom iterative CORDIC arithmetic, precision-switching behaviour, and execution scheduling. Fixed-point arithmetic is implemented using the FxP-Math library, while neural network layers and quantised inference flows are modelled using QKeras~2.3.

The software framework supports configurable precision modes (8-bit and 16-bit), variable CORDIC iteration depth, and layer-wise execution control. All deep learning evaluations are performed against an FP32 reference baseline under identical network topology, dataset, and inference conditions. This approach ensures that observed accuracy differences are attributable solely to arithmetic approximation, not to changes in training or model structure.

Accuracy is evaluated at both the layer and end-to-end model levels for representative CNN and transformer-style MLP workloads. The number of CORDIC iterations per layer is selected using an accuracy-sensitivity heuristic~\cite{Flex-PE}, which identifies numerically critical layers and assigns them to accurate execution modes, while non-critical layers operate in approximate mode.

\subsection{RTL Modelling and Functional Verification}

The proposed iterative CORDIC-based MAC unit and vector-engine datapath are modelled in synthesizable Verilog HDL. The architecture is parameterised to support different vector widths, precision modes, and iteration depths. A cycle-accurate RTL testbench is developed to validate functional correctness across all supported operating modes.

Functional verification is performed using Synopsys VCS, where RTL outputs are compared against the software emulation model for a wide range of randomised and application-driven test vectors. This cross-validation ensures bit-level consistency between the software model and the hardware implementation, accounting for fixed-point rounding, truncation, and iteration control.

\subsection{FPGA Prototyping and Measurement}

FPGA-based evaluation is conducted using the AMD Virtex-707 (VC707) platform. Synthesis, placement, and routing are performed using the AMD Vivado Design Suite with a target operating frequency of 100\, MHz. All reported FPGA metrics, including lookup tables (LUTs), flip-flops (FFs), timing, and power consumption, are obtained from post-place-and-route reports to avoid optimistic estimation.

To enable fair comparison with state-of-the-art designs, either the reported post-implementation results from prior work are used directly, or the designs are re-synthesised under comparable constraints where feasible. Power measurements are extracted using vendor-supported power analysis tools with realistic switching activity derived from application traces.

\subsection{ASIC Synthesis and Technology Assumptions}

ASIC evaluation is carried out using Synopsys Design Compiler targeting a commercial 28\, nm HPC+ CMOS technology at 0.9\, V. Standard-cell libraries for worst-case timing corners are used to ensure conservative delay estimates. Area, timing, and power metrics are extracted from post-synthesis reports.

System-level performance metrics, including energy efficiency (TOPS/W) and compute density (TOPS/mm\textsuperscript{2}), are derived using consistent workload assumptions across all designs. The same precision mode, vector width, and clock frequency normalisation are applied when comparing against prior accelerators to ensure fairness.

\subsection{System-Level Deployment and End-to-End Validation}

To validate practical applicability, the proposed vector engine is deployed on a Pynq-Z2 platform with an ARM Cortex-A9 host processor. The accelerator is integrated through an AXI-based interface and evaluated on object detection and classification workloads. End-to-end latency and power consumption are measured at the application level, capturing the combined effects of computation, data movement, and control overhead.

This multi-level evaluation methodology ensures that the reported improvements are not limited to isolated circuit optimisations but translate into tangible system-level benefits for real-world edge AI deployments.

\section{Results and Discussion}

This section presents a comprehensive evaluation of the proposed CORDIC-based vector engine at the circuit, architectural, and system levels. Results are reported for FPGA prototyping, ASIC synthesis, and end-to-end deployment and are compared against representative state-of-the-art (SoTA) AI accelerators to highlight performance, energy efficiency, and scalability trade-offs.

\begin{table*}[!t]
\caption{Comparative Performance Metrics for CORDIC-based different SoTA  MAC units}
\label{tab:mac-util-comp}
\renewcommand{\arraystretch}{1.15}
\resizebox{\textwidth}{!}{%
\begin{tabular}{|cccccccccccccc|}
\hline
\multicolumn{1}{|c|}{\textbf{Design}} & \multicolumn{2}{c|}{\textbf{TCAS-II'24\cite{RPE}}} & \multicolumn{2}{c|}{\textbf{ISCAS'25\cite{LPRE}}} & \multicolumn{5}{c|}{\textbf{ICIIS'25\cite{HYDRA_ICIIS25}}} & \multicolumn{1}{c|}{\textbf{TVLSI'25\cite{MSDF-MAC}}} & \multicolumn{1}{c|}{\textbf{TCAD'22\cite{Acc-App-MAC}}} & \multicolumn{1}{c|}{\textbf{TVLSI'25\cite{Flex-PE}}} & \textbf{Proposed} \\ \hline
\multicolumn{14}{|c|}{{\textbf{FPGA Utilization (VC707, 100 MHz)}}} \\ \hline
\multicolumn{1}{|c|}{\textbf{Parameter}} & \multicolumn{1}{c|}{\textbf{FP32}} & \multicolumn{1}{c|}{\textbf{FP32}} & \multicolumn{1}{c|}{\textbf{BF16}} & \multicolumn{1}{c|}{\textbf{Posit-8}} & \multicolumn{1}{c|}{\textbf{Vedic}} & \multicolumn{1}{c|}{\textbf{Wallace}} & \multicolumn{1}{c|}{\textbf{Booth}} & \multicolumn{1}{c|}{\textbf{Quant-MAC}} & \multicolumn{1}{c|}{\textbf{CORDIC}} & \multicolumn{1}{c|}{\textbf{MSDF-MAC}} & \multicolumn{1}{c|}{\textbf{Acc-App-MAC}} & \multicolumn{1}{c|}{\textbf{CORDIC}} & \textbf{Iter-MAC} \\ \hline
\multicolumn{1}{|c|}{\textbf{LUTs}} & \multicolumn{1}{c|}{8065} & \multicolumn{1}{c|}{8054} & \multicolumn{1}{c|}{3670} & \multicolumn{1}{c|}{467} & \multicolumn{1}{c|}{160} & \multicolumn{1}{c|}{106} & \multicolumn{1}{c|}{84} & \multicolumn{1}{c|}{72} & \multicolumn{1}{c|}{56} & \multicolumn{1}{c|}{62} & \multicolumn{1}{c|}{57} & \multicolumn{1}{c|}{45} & 24 \\ \hline
\multicolumn{1}{|c|}{\textbf{FFs}} & \multicolumn{1}{c|}{1072} & \multicolumn{1}{c|}{1718} & \multicolumn{1}{c|}{324} & \multicolumn{1}{c|}{175} & \multicolumn{1}{c|}{241} & \multicolumn{1}{c|}{113} & \multicolumn{1}{c|}{59} & \multicolumn{1}{c|}{56} & \multicolumn{1}{c|}{72} & \multicolumn{1}{c|}{45} & \multicolumn{1}{c|}{NR} & \multicolumn{1}{c|}{37} & 22 \\ \hline
\multicolumn{1}{|c|}{\textbf{Delay (ns)}} & \multicolumn{1}{c|}{5.56} & \multicolumn{1}{c|}{4.6} & \multicolumn{1}{c|}{0.512} & \multicolumn{1}{c|}{2.68} & \multicolumn{1}{c|}{4.5} & \multicolumn{1}{c|}{2.6} & \multicolumn{1}{c|}{3.1} & \multicolumn{1}{c|}{5.4} & \multicolumn{1}{c|}{1.52} & \multicolumn{1}{c|}{3.2} & \multicolumn{1}{c|}{3.51} & \multicolumn{1}{c|}{4.5} & 9.1 \\ \hline
\multicolumn{1}{|c|}{\textbf{Power (mW)}} & \multicolumn{1}{c|}{378} & \multicolumn{1}{c|}{296} & \multicolumn{1}{c|}{136} & \multicolumn{1}{c|}{68} & \multicolumn{1}{c|}{6.1} & \multicolumn{1}{c|}{3.3} & \multicolumn{1}{c|}{3.1} & \multicolumn{1}{c|}{4.2} & \multicolumn{1}{c|}{8.3} & \multicolumn{1}{c|}{5.8} & \multicolumn{1}{c|}{6.9} & \multicolumn{1}{c|}{2} & 1.9 \\ \hline
\multicolumn{1}{|c|}{\textbf{PDP (pJ)}} & \multicolumn{1}{c|}{2102} & \multicolumn{1}{c|}{1361.6} & \multicolumn{1}{c|}{69.6} & \multicolumn{1}{c|}{182} & \multicolumn{1}{c|}{27.45} & \multicolumn{1}{c|}{8.58} & \multicolumn{1}{c|}{9.6} & \multicolumn{1}{c|}{22.68} & \multicolumn{1}{c|}{12.6} & \multicolumn{1}{c|}{18.56} & \multicolumn{1}{c|}{24.2} & \multicolumn{1}{c|}{9} & 17.29 \\ \hline
\multicolumn{14}{|c|}{{\textbf{ASIC Performance (28nm, 0.9V)}}} \\ \hline
\multicolumn{1}{|c|}{\textbf{Area (um\textasciicircum{}2)}} & \multicolumn{1}{c|}{10000} & \multicolumn{1}{c|}{13000} & \multicolumn{1}{c|}{4340} & \multicolumn{1}{c|}{754} & \multicolumn{1}{c|}{407} & \multicolumn{1}{c|}{296} & \multicolumn{1}{c|}{271} & \multicolumn{1}{c|}{175} & \multicolumn{1}{c|}{264} & \multicolumn{1}{c|}{286} & \multicolumn{1}{c|}{259} & \multicolumn{1}{c|}{8570} & 108 \\ \hline
\multicolumn{1}{|c|}{\textbf{Delay (ns)}} & \multicolumn{1}{c|}{679} & \multicolumn{1}{c|}{700} & \multicolumn{1}{c|}{295} & \multicolumn{1}{c|}{40.6} & \multicolumn{1}{c|}{6.38} & \multicolumn{1}{c|}{5.62} & \multicolumn{1}{c|}{5.3} & \multicolumn{1}{c|}{3.58} & \multicolumn{1}{c|}{2.36} & \multicolumn{1}{c|}{1.42} & \multicolumn{1}{c|}{2.6} & \multicolumn{1}{c|}{0.7} & 2.98 \\ \hline
\multicolumn{1}{|c|}{\textbf{Power (mW)}} & \multicolumn{1}{c|}{15.86} & \multicolumn{1}{c|}{29.3} & \multicolumn{1}{c|}{6.89} & \multicolumn{1}{c|}{1.8} & \multicolumn{1}{c|}{35} & \multicolumn{1}{c|}{37} & \multicolumn{1}{c|}{12.8} & \multicolumn{1}{c|}{89} & \multicolumn{1}{c|}{24.5} & \multicolumn{1}{c|}{6.7} & \multicolumn{1}{c|}{12.4} & \multicolumn{1}{c|}{1.5} & 6.3 \\ \hline
\multicolumn{1}{|c|}{\textbf{PDP (pJ)}} & \multicolumn{1}{c|}{10768.94} & \multicolumn{1}{c|}{20510} & \multicolumn{1}{c|}{4682} & \multicolumn{1}{c|}{1189} & \multicolumn{1}{c|}{223.3} & \multicolumn{1}{c|}{207.94} & \multicolumn{1}{c|}{67.84} & \multicolumn{1}{c|}{318.62} & \multicolumn{1}{c|}{57.82} & \multicolumn{1}{c|}{9.514} & \multicolumn{1}{c|}{32.24} & \multicolumn{1}{c|}{1.05} & 18.774 \\ \hline
\end{tabular}}
\end{table*}

\begin{table*}[!t]
\caption{Comparative Performance Metrics for CORDIC-based different SoTA  AF units}
\label{tab:resource-af}
\renewcommand{\arraystretch}{1.45}
\resizebox{\textwidth}{!}{%
\begin{tabular}{|ccccccccccccccc|}
\hline
\multicolumn{1}{|c|}{\textbf{Design}} & \multicolumn{3}{c|}{\textbf{ISQED'24\cite{FP-CORDIC-ISQED24}}} & \multicolumn{1}{c|}{\textbf{TCAS-II'20\cite{Precision-Scalable-AF-TCAS-II20}}} & \multicolumn{1}{c|}{\textbf{TVLSI'23\cite{AxAF-TVLSI23}}} & \multicolumn{3}{c|}{\textbf{ISQED'24\cite{FP-CORDIC-ISQED24}}} & \multicolumn{1}{c|}{\textbf{TC'23\cite{LSTM-AF-TC}}} & \multicolumn{3}{c|}{\textbf{ISQED'24\cite{FP-CORDIC-ISQED24}}} & \multicolumn{1}{c|}{\textbf{TVLSI'25\cite{Flex-PE}}} & \textbf{Proposed} \\ \hline
\multicolumn{15}{|c|}{{\textbf{FPGA Utilization (VC707, 100 MHz)}}} \\ \hline
\multicolumn{1}{|c|}{\textbf{Parameter}} & \multicolumn{1}{c|}{Softmax-FP32} & \multicolumn{1}{c|}{Softmax-FP16} & \multicolumn{1}{c|}{Softmax-BF16} & \multicolumn{1}{c|}{Softmax- FxP8/16} & \multicolumn{1}{c|}{Softmax-16b} & \multicolumn{1}{c|}{Tanh-FP32} & \multicolumn{1}{c|}{Tanh-FP16} & \multicolumn{1}{c|}{Tanh-BF16} & \multicolumn{1}{c|}{Tanh/Sigmoid-16b} & \multicolumn{1}{c|}{Sigmoid-FP32} & \multicolumn{1}{c|}{Sigmoid-FP16} & \multicolumn{1}{c|}{Sigmoid-BF16} & \multicolumn{1}{c|}{SSTp} & FxP-4/8/16 \\ \hline
\multicolumn{1}{|c|}{\textbf{LUTs}} & \multicolumn{1}{c|}{3217} & \multicolumn{1}{c|}{1137} & \multicolumn{1}{c|}{1263} & \multicolumn{1}{c|}{2564} & \multicolumn{1}{c|}{1215} & \multicolumn{1}{c|}{4298} & \multicolumn{1}{c|}{1530} & \multicolumn{1}{c|}{1513} & \multicolumn{1}{c|}{2395} & \multicolumn{1}{c|}{5101} & \multicolumn{1}{c|}{1853} & \multicolumn{1}{c|}{1856} & \multicolumn{1}{c|}{897} & 537 \\ \hline
\multicolumn{1}{|c|}{\textbf{FFs}} & \multicolumn{1}{c|}{NR} & \multicolumn{1}{c|}{NR} & \multicolumn{1}{c|}{NR} & \multicolumn{1}{c|}{2794} & \multicolumn{1}{c|}{1012} & \multicolumn{1}{c|}{NR} & \multicolumn{1}{c|}{NR} & \multicolumn{1}{c|}{NR} & \multicolumn{1}{c|}{1503} & \multicolumn{1}{c|}{NR} & \multicolumn{1}{c|}{NR} & \multicolumn{1}{c|}{NR} & \multicolumn{1}{c|}{1231} & 468 \\ \hline
\multicolumn{1}{|c|}{\textbf{Delay (ns)}} & \multicolumn{1}{c|}{92} & \multicolumn{1}{c|}{43} & \multicolumn{1}{c|}{45} & \multicolumn{1}{c|}{2.3} & \multicolumn{1}{c|}{3.32} & \multicolumn{1}{c|}{56} & \multicolumn{1}{c|}{34} & \multicolumn{1}{c|}{38} & \multicolumn{1}{c|}{0.18} & \multicolumn{1}{c|}{109} & \multicolumn{1}{c|}{60} & \multicolumn{1}{c|}{45} & \multicolumn{1}{c|}{11.8} & 2.6 \\ \hline
\multicolumn{1}{|c|}{\textbf{Power (mW)}} & \multicolumn{1}{c|}{115} & \multicolumn{1}{c|}{115} & \multicolumn{1}{c|}{77} & \multicolumn{1}{c|}{NR} & \multicolumn{1}{c|}{165} & \multicolumn{1}{c|}{130} & \multicolumn{1}{c|}{124} & \multicolumn{1}{c|}{82} & \multicolumn{1}{c|}{681} & \multicolumn{1}{c|}{121} & \multicolumn{1}{c|}{118} & \multicolumn{1}{c|}{83} & \multicolumn{1}{c|}{59} & 30 \\ \hline
\multicolumn{1}{|c|}{\textbf{PDP (pJ)}} & \multicolumn{1}{c|}{10580} & \multicolumn{1}{c|}{4945} & \multicolumn{1}{c|}{3465} & \multicolumn{1}{c|}{-} & \multicolumn{1}{c|}{548} & \multicolumn{1}{c|}{7280} & \multicolumn{1}{c|}{4216} & \multicolumn{1}{c|}{3116} & \multicolumn{1}{c|}{123} & \multicolumn{1}{c|}{13189} & \multicolumn{1}{c|}{7080} & \multicolumn{1}{c|}{3735} & \multicolumn{1}{c|}{696.2} & 78 \\ \hline
\multicolumn{15}{|c|}{{ \textbf{ASIC Performance (28nm, 0.9V)}}} \\ \hline
\multicolumn{1}{|c|}{\textbf{Area (um\textasciicircum{}2)}} & \multicolumn{1}{c|}{41536} & \multicolumn{1}{c|}{17289} & \multicolumn{1}{c|}{11301} & \multicolumn{1}{c|}{18392} & \multicolumn{1}{c|}{3819} & \multicolumn{1}{c|}{5060} & \multicolumn{1}{c|}{1180} & \multicolumn{1}{c|}{843} & \multicolumn{1}{c|}{870523} & \multicolumn{1}{c|}{2234} & \multicolumn{1}{c|}{1855} & \multicolumn{1}{c|}{1180} & \multicolumn{1}{c|}{49152} & 2138 \\ \hline
\multicolumn{1}{|c|}{\textbf{Delay (ns)}} & \multicolumn{1}{c|}{6} & \multicolumn{1}{c|}{4} & \multicolumn{1}{c|}{3.3} & \multicolumn{1}{c|}{0.3} & \multicolumn{1}{c|}{1.6} & \multicolumn{1}{c|}{4} & \multicolumn{1}{c|}{3.3} & \multicolumn{1}{c|}{3.4} & \multicolumn{1}{c|}{NR} & \multicolumn{1}{c|}{7.6} & \multicolumn{1}{c|}{4.4} & \multicolumn{1}{c|}{3.26} & \multicolumn{1}{c|}{2.3} & 2.6 \\ \hline
\multicolumn{1}{|c|}{\textbf{Power (mW)}} & \multicolumn{1}{c|}{75} & \multicolumn{1}{c|}{40} & \multicolumn{1}{c|}{25} & \multicolumn{1}{c|}{51.6} & \multicolumn{1}{c|}{1.6} & \multicolumn{1}{c|}{8.75} & \multicolumn{1}{c|}{3} & \multicolumn{1}{c|}{2} & \multicolumn{1}{c|}{150} & \multicolumn{1}{c|}{10} & \multicolumn{1}{c|}{4.8} & \multicolumn{1}{c|}{2.5} & \multicolumn{1}{c|}{5.2} & 60 \\ \hline
\multicolumn{1}{|c|}{\textbf{PDP (pJ)}} & \multicolumn{1}{c|}{450} & \multicolumn{1}{c|}{160} & \multicolumn{1}{c|}{82.5} & \multicolumn{1}{c|}{15.5} & \multicolumn{1}{c|}{2.56} & \multicolumn{1}{c|}{35} & \multicolumn{1}{c|}{9.9} & \multicolumn{1}{c|}{6.8} & \multicolumn{1}{c|}{-} & \multicolumn{1}{c|}{76} & \multicolumn{1}{c|}{21.12} & \multicolumn{1}{c|}{8.15} & \multicolumn{1}{c|}{11.96} & 156 \\ \hline
\end{tabular}}
\end{table*}

\subsection{MAC-Level Hardware Efficiency}

Table~\ref{tab:mac-util-comp} compares the proposed iterative CORDIC-based MAC unit with prior CORDIC, logarithmic, and approximation-based MAC designs across both FPGA and ASIC platforms. On the Virtex-707 FPGA, the proposed MAC achieves significant reductions in lookup tables (LUTs) and flip-flops (FFs) compared to pipelined CORDIC and fixed-point MAC designs, while avoiding the use of DSP blocks. This reduction directly translates into lower static power consumption and improved placement flexibility.

At the ASIC level (28\, nm, 0.9\, V), the proposed MAC demonstrates up to 33\% reduction in critical-path delay and approximately 21\% lower power per MAC stage compared to comparable CORDIC-based designs. Although the iterative MAC incurs a multi-cycle execution latency, this overhead is amortised at the vector-engine level through parallel execution across multiple processing elements (PEs), as discussed in Section~II. Consequently, the proposed design achieves a favourable power-delay product (PDP) while maintaining runtime configurability between approximate and accurate execution modes.

\subsection{Activation-Function Hardware Utilization}

Table~\ref{tab:resource-af} summarises the FPGA and ASIC resource utilisation of the proposed time-multiplexed multi-activation-function (multi-AF) block relative to prior dedicated AF implementations. Existing designs often allocate separate hardware blocks for individual activation functions\cite{TC5_AF}, leading to significant underutilization and dark silicon. In contrast, the proposed multi-AF block reuses CORDIC resources across multiple nonlinear functions, including Sigmoid, Tanh, SoftMax, GELU, Swish, ReLU, and SELU.

The results show that the proposed design achieves utilisation factors of 72-86\% depending on the activation mode, while incurring less than 4\% additional area and power overhead. On an FPGA, the multi-AF block reduces LUT and FF usage compared to SoTA designs supporting a similar function set. On ASIC, it demonstrates lower power consumption and competitive delay, confirming that time multiplexing effectively mitigates activation-function underutilization without compromising performance.

\subsection{Accuracy Evaluation Under Iterative Approximation}

\begin{figure}[!t]
    \centering
    \includegraphics[width=\columnwidth]{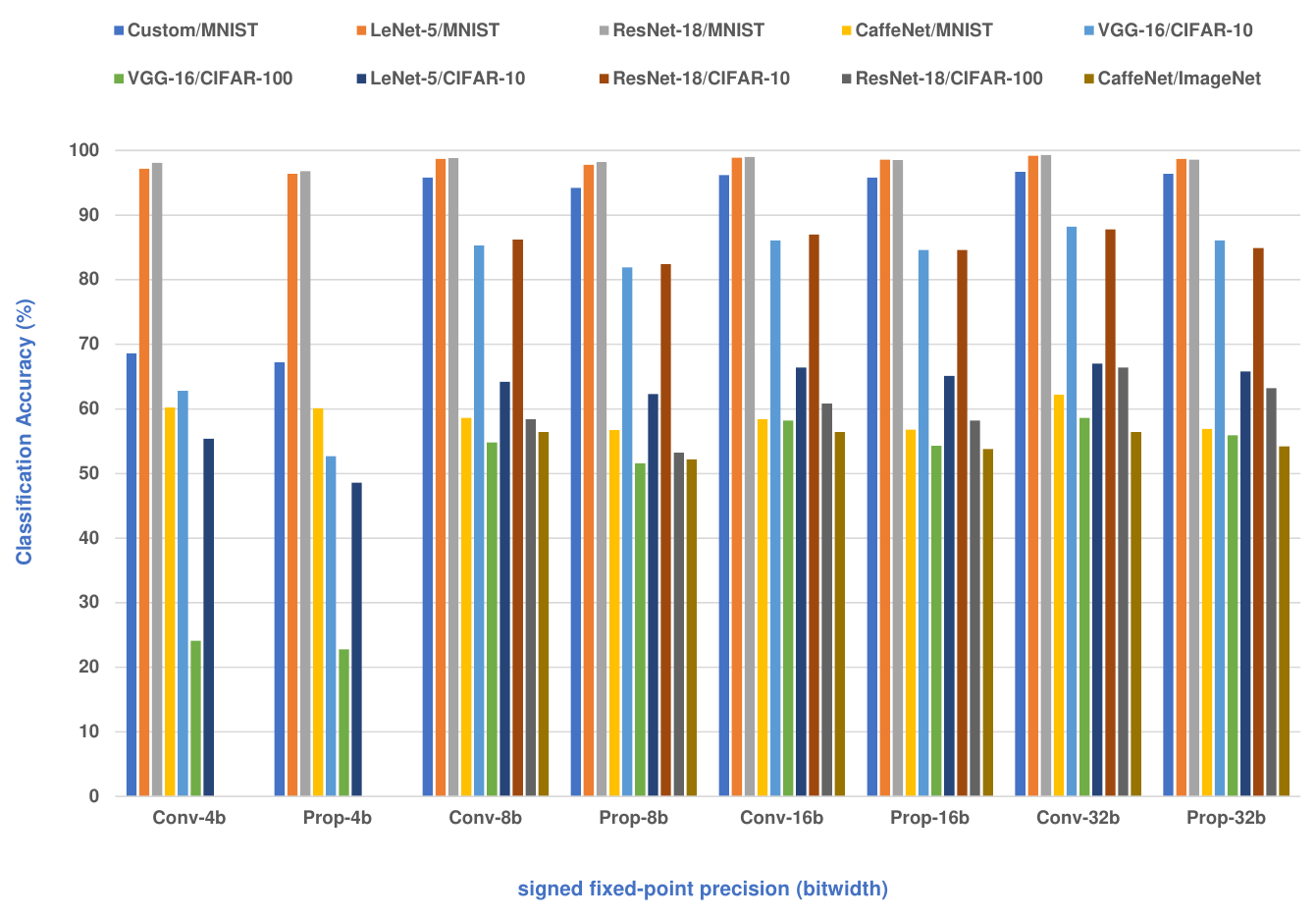}
    \caption{Evaluation of DNN accuracy for different DNN models with CORDIC methodology.}
    \label{fig:accuracy}
\end{figure}

Fig.~\ref{fig:accuracy} reports the accuracy of representative CNN and DNN models under different CORDIC iteration settings. The results confirm that numerical error is tightly coupled to the number of active CORDIC iterations, validating the effectiveness of the proposed runtime accuracy-latency trade-off mechanism. When operating in approximate mode, the accelerator incurs approximately 2\% accuracy degradation, while accurate mode limits accuracy loss to below 0.5\%.

Importantly, by applying an accuracy-sensitivity heuristic to select the iteration depth per layer, most of the performance benefits of approximate execution are retained while preserving end-to-end model accuracy. This demonstrates that the proposed architecture enables fine-grained control over numerical fidelity without requiring retraining or auxiliary correction hardware.

\subsection{FPGA System-Level Comparison}

Table~\ref{tab:arch-fpga} compares the proposed vector engine against SoTA FPGA-based AI accelerators using object detection workloads such as TinyYOLO-v3. The proposed design achieves competitive throughput while significantly reducing power consumption. Operating at 85.4\, MHz on the Virtex-707 platform, the vector engine delivers 6.43\, GOPS/W at only 0.53\, W, outperforming several prior designs in energy efficiency despite using no DSP blocks.

Compared to designs such as Flex-PE and LPRE, which rely on higher operating frequencies or specialised arithmetic units, the proposed architecture emphasises energy efficiency and scalability, making it particularly well-suited for edge deployments where power budgets are tightly constrained.

\subsection{ASIC Scalability and Compute Density}

ASIC-level scalability is evaluated using two configurations of the proposed vector engine: a 64-PE configuration and a 256-PE configuration, as reported in Table~\ref{tab:arch-asic}. The 64-PE configuration serves as a computationally equivalent baseline, demonstrating comparable performance to prior designs at significantly lower area and power. The 256-PE configuration represents a resource-equivalent comparison, achieving a peak compute density of 4.83~TOPS/mm\textsuperscript{2} and an energy efficiency of 11.67~TOPS/W.

These results highlight the benefits of the proposed iterative execution model, where increased vector width compensates for per-MAC latency while preserving energy efficiency. The architecture's scalability enables efficient deployment across a wide range of performance targets without redesigning the core datapath.

\begin{table*}[!t]
\caption{Analysis of FPGA Hardware Implementation for object detection (TinyYolo-v3) with SoTA AI accelerator designs}
\label{tab:arch-fpga}
\renewcommand{\arraystretch}{1.05}
\resizebox{\textwidth}{!}{%
\begin{tabular}{|c|c|c|c|c|c|c|c|c|}
\hline
\textbf{Design} & \textbf{Platform} & \textbf{Precision} & \textbf{k-LUTs} & \textbf{k-Regs/FFs} & \textbf{DSPs} & \textbf{\begin{tabular}[c]{@{}c@{}}Op. Freq \\      (MHz)\end{tabular}} & \textbf{\begin{tabular}[c]{@{}c@{}}Energy efficiency \\      (GOPS/W)\end{tabular}} & \textbf{Power(W)} \\ \hline
\textbf{Proposed} & VC707 & 4/8/16 & 26.7 & 15.9 & - & 85.4 & 6.43 & 0.53 \\ \hline
\textbf{TVLSI'25\cite{Flex-PE}} & VC707 & 4/8/16/32 & 38.7 & 17.4 & 73 & 466 & 8.42 & 2.24 \\ \hline
\textbf{TCAS-I'24\cite{BWu-TCASI24}} & ZU3EG & 8 & 40.8 & 45.5 & 258 & 100 & 0.39 & 2.2 \\ \hline
\textbf{TCAS-II'23\cite{Ski-TCASII23}} & XCVU9P & 8 & 132 & 39.5 & 96 & 150 & 6.36 & 5.52 \\ \hline
\textbf{TVLSI'23\cite{lee-TVLSI}} & ZCU102 & 8 & 117 & 74 & 132 & 300 & 4.2 & 6.58 \\ \hline
\textbf{Access'24\cite{QuantMAC}} & VC707 & 4/8 & 19.8 & 12.1 & 39 & 136 & 0.68 & 1.81 \\ \hline
\textbf{ISCAS'25\cite{LPRE}} & VCU129 & 8/16/32 & 17.5 & 14.8 & - & 54.5 & 2.64 & 1.6 \\ \hline
\end{tabular}}
\end{table*}

\subsection{End-to-End Embedded Deployment}

\begin{figure}[!t]
     \centering
     \includegraphics[width=0.975\columnwidth]{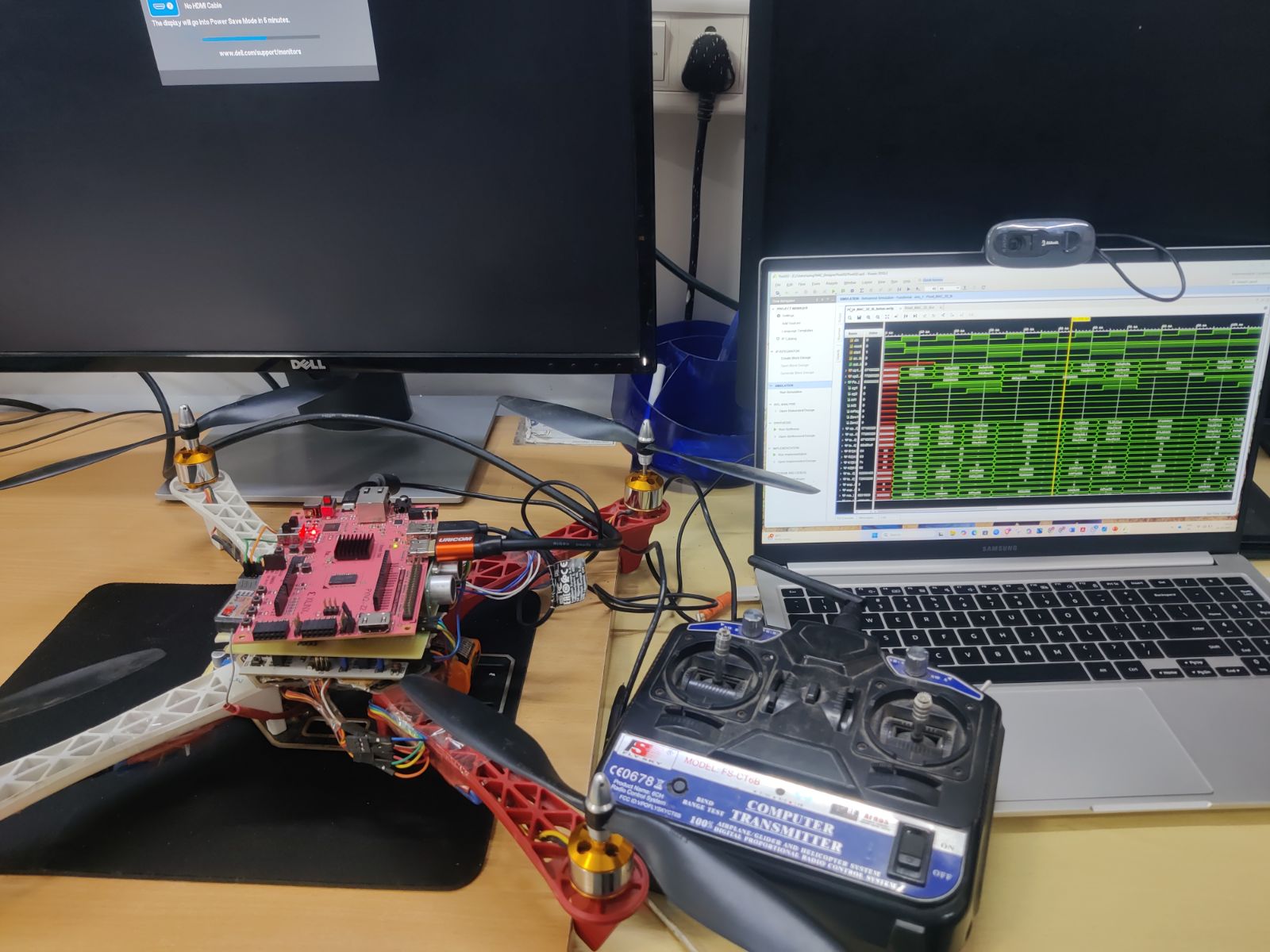}
     \caption{Prototype visualisation, showing Pynq-z2 for edge AI inference on Unmanned Aerial Vehicles (UAV).}
     \label{visualization}
\end{figure}

\begin{figure}[!t]
    \centering
    \includegraphics[width=\columnwidth]{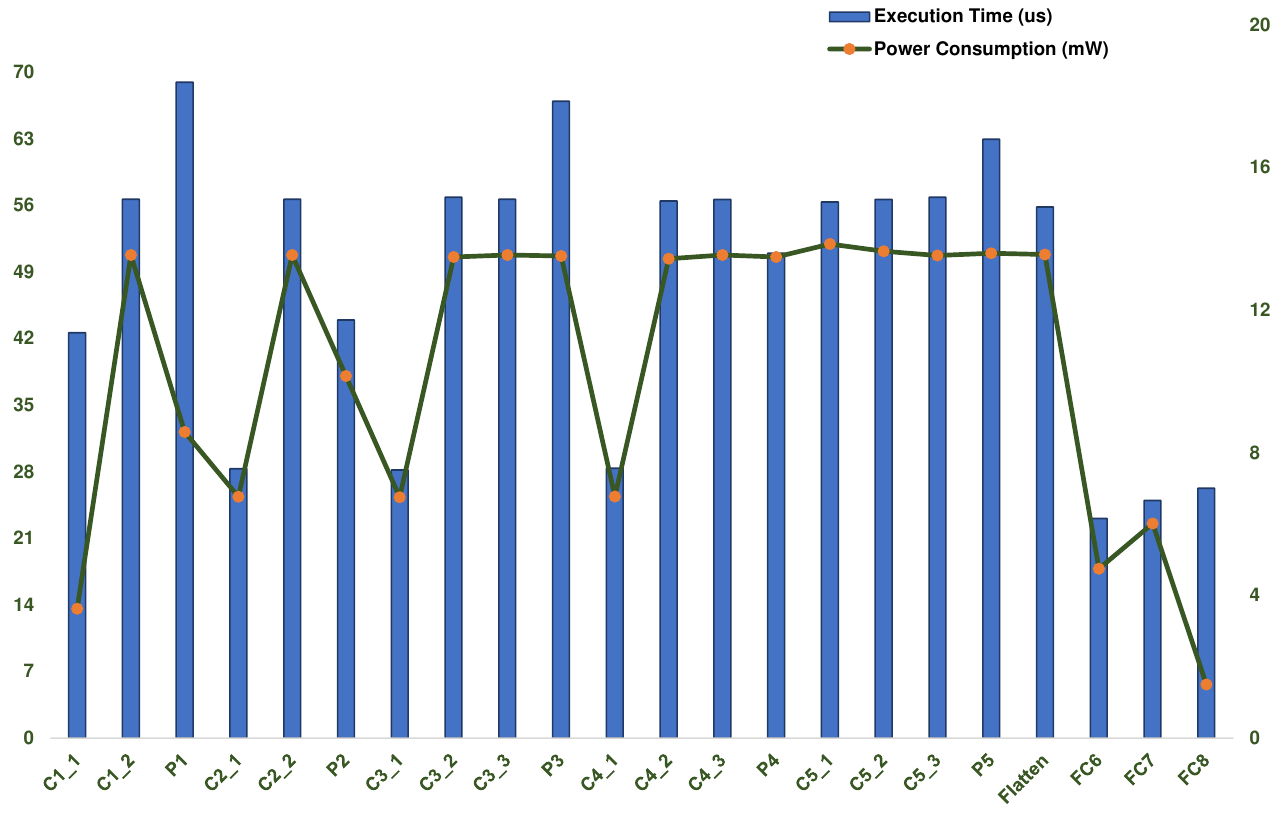}
    \caption{VGG-16 layer-wise execution time and power consumption.}
    \label{fig:vgg16_perf}
\end{figure}

Fig.~\ref{fig:vgg16_perf} presents a layer-wise execution-time and power breakdown for the VGG-16 model, illustrating the impact of runtime precision switching on system performance. End-to-end deployment on a Pynq-Z2 platform with an ARM Cortex-A9 host reports a total latency of 84.6\, ms at 0.43\, W for object detection and classification workloads, as shown in Fig. \ref{visualization}. It outperforms prior works: \cite{Flex-PE} ($186.4 ms$ / $2.24$ W on $VC707$), \cite{GR-ACMTr} ($772$ ms / $1.524$ W on $VC707$), \cite{LPRE} ($184$ ms / $0.93$ W on Pynq-Z$2$), \cite{access_ratko} ($163.7$ ms / $13.32$ W on $VCU102$), and baselines such as NVIDIA Jetson Nano ($226$ ms / $1.34$ W) and Raspberry Pi ($555$ ms / $2.7$ W). 

The proposed vector engine achieves lower latency and lower power consumption than prior FPGA-based accelerators and commercial embedded platforms such as the NVIDIA Jetson Nano and Raspberry Pi. These improvements stem from a combination of iterative MAC efficiency, reduced memory bandwidth requirements, and dynamic precision adaptation.

Overall, the results demonstrate that the proposed architecture delivers consistent improvements across circuit-level efficiency, architectural scalability, and system-level performance, validating its suitability for energy-efficient edge AI acceleration.

\begin{table*}[!t]
\caption{ASIC performance comparison with SoTA 8-bit accelerator designs, with CMOS 28nm, 0.9V, SF technology.}
\label{tab:arch-asic}
\renewcommand{\arraystretch}{1.2}
\resizebox{\textwidth}{!}{%
\begin{tabular}{|l|c|l|c|c|c|c|c|}
\hline
\textbf{Design} & \textbf{Network/Arch} & \textbf{Datatype} & \textbf{Freq. (GHz)} & \textbf{Area (mm\textsuperscript{2})} & \textbf{Power (mW)} & \begin{tabular}[c]{@{}c@{}}\textbf{Energy Efficiency}\\ \textbf{TOPS/W}\end{tabular} & \begin{tabular}[c]{@{}c@{}}\textbf{Compute Density}\\ TOPS/mm\textsuperscript{2}\end{tabular} \\ \hline

\multirow{2}{*}{\textbf{TCAS-II'24\cite{RPE}}} & \multirow{2}{*}{Vector Engine (64$\times$MACs)} & \multirow{2}{*}{FP8} & 1.47 & 0.896 & 1622 & 7.24 & 2.39 \\ \cline{4-8} 
 &  &  & 1.29 & 1.18 & 1375 & 3.57 & 1.21 \\ \hline
 \textbf{TCAS-I'22\cite{ILM}} & \begin{tabular}[c]{@{}c@{}}Vector Engine (64$\times$MACs)\\ 196-64-32-32-10\end{tabular} & INT-8 & 0.4 & 2.43 & 224.6 & 7.75 & 1.67 \\ \hline
\textbf{ISCAS'25\cite{LPRE}} & \begin{tabular}[c]{@{}c@{}}TREA (64$\times$MACs)\\ 196-64-32-32-10\end{tabular} & Posit-8 & 1.25 & 6.73 & 230.4 & 7.55 & 0.16 \\ \hline
 \textbf{TVLSI'25\cite{Flex-PE}} & Systolic Array (8x8) & FxP8& 0.44 & 1.85 & 523 & 4.3 & 2.76 \\ \hline
\textbf{ICIIS'25\cite{HYDRA_ICIIS25}} & \begin{tabular}[c]{@{}c@{}}Layer-Reused (64$\times$MACs)\\ 196-64-32-32-10\end{tabular} & FxP8 & 0.25 & 3.78 & 1540 & 4.28 & 2.07 \\ \hline
\multirow{2}{*}{\textbf{Proposed}} & Vector Engine 64$\times$PEs & \multirow{2}{*}{FxP-4/8/16} & 1.24 & 0.43 & 329 & 3.84 & 1.52 \\ \cline{2-2} \cline{4-8} 
 & Vector Engine 256$\times$PEs &  & 0.96 & 1.42 & 1186 & 11.67 & 4.83 \\ \hline
 \textbf{Access'24\cite{QuantMAC}} & \begin{tabular}[c]{@{}c@{}}Shared Bank (256$\times$MACs)\\ 784-196-120-84-10\end{tabular} & FxP8 & 0.28 & 1.58 & 499.7 & 6.87 & 1.18 \\ \hline
\end{tabular}}
\end{table*}

\section{Conclusion and Future Work}

This paper presented a runtime-adaptive, CORDIC-accelerated vector engine designed to address the efficiency and flexibility challenges of deep learning inference on resource-constrained edge platforms. By introducing a low-resource, iterative CORDIC-based MAC unit with runtime-configurable iteration depth, the proposed architecture enables fine-grained trade-offs between accuracy and latency without requiring auxiliary error-correction hardware or structural modifications. In contrast to prior fixed-approximation designs, this approach allows dynamic adaptation to layer-level numerical sensitivity while maintaining compatibility with standard deep learning workloads.

The proposed vector engine further integrates a time-multiplexed multi-activation-function (multi-AF) block that significantly improves hardware utilisation and mitigates dark silicon. By sharing CORDIC resources across a wide range of nonlinear functions, including Sigmoid, Tanh, SoftMax, GELU, Swish, ReLU, and SELU, the architecture achieves high utilisation factors with minimal additional area and power overhead. This balanced treatment of MAC and activation units addresses a persistent inefficiency in existing deep learning accelerators.

Comprehensive evaluation across software emulation, RTL verification, FPGA prototyping, and 28\, nm ASIC synthesis demonstrates the effectiveness of the proposed design. The iterative MAC unit achieves up to 33\% reduction in critical-path delay and 21\% power savings per stage, while scalable vector-engine configurations deliver a peak compute density of 4.83~TOPS/mm\textsuperscript{2} and an energy efficiency of 11.67~TOPS/W. End-to-end deployment on embedded platforms further confirms that the architectural benefits translate into tangible improvements in latency and power consumption at the system level.

Future work will focus on extending the proposed framework toward a compiler-assisted design flow that automates layer-wise precision and iteration selection based on model sensitivity analysis. In addition, integrating full physical design and place-and-route (PnR) optimisation will enable more accurate post-layout evaluation and facilitate tape-out readiness. Further exploration of adaptive execution strategies for emerging transformer and multi-modal workloads, as well as tighter integration with RISC-V-based system-on-chip platforms, represents promising directions for expanding the applicability of the proposed vector engine\cite{TC2_riscv}.

Overall, the proposed runtime-adaptive CORDIC-based vector engine provides a scalable and energy-efficient compute substrate that bridges the gap between approximate and accurate deep learning acceleration, making it well-suited for next-generation edge AI systems.

\bibliographystyle{ieeetr}
\bibliography{this_bibliography}

\end{document}